\documentclass[twocolumn,prl,floatfix,superscriptaddress]{revtex4}

\usepackage{graphicx}
\usepackage{amsmath}
\usepackage{latexsym}
\usepackage{epstopdf}
\usepackage{amsmath}

\newcommand{\beq}{\begin{equation}}
\newcommand{\eeq}{\end{equation}}

\begin{document}

\title{Tunable bandgaps and excitons in doped semiconducting carbon
nanotubes made possible by acoustic plasmons}
\author{Catalin D. Spataru$^*$}
\affiliation{Sandia National Laboratories, Livermore, CA 94551, USA}
\author{Fran\c{c}ois L\'{e}onard}
\affiliation{Sandia National Laboratories, Livermore, CA 94551, USA}

\begin{abstract}
Doping of semiconductors is essential in modern electronic and photonic
devices. While doping is well understood in bulk semiconductors, the advent
of carbon nanotubes and nanowires for nanoelectronic and nanophotonic
applications raises some key questions about the role and impact of doping
at low dimensionality. Here we show that for semiconducting carbon
nanotubes, bandgaps and exciton binding energies can be dramatically reduced
upon experimentally relevant doping, and can be tuned gradually over a broad
range of energies in contrast to higher dimensional systems. The later
feature is made possible by a novel mechanism involving strong dynamical
screening effects mediated by acoustic plasmons.
\end{abstract}

\maketitle

Nanomaterials have been lauded for their promise in electronic and photonic
applications \cite{Avouris}. Quite often, the imagined nanodevices rely on
analogies with those based on bulk semiconductors. However, the true
potential of nanomaterials lies in the exploitation of their unique
properties to realize entirely new device concepts. In particular,
approaches for externally controlling their electronic and optical properties 
would enable new strategies for device design. 

Here we propose that such control is possible in carbon nanotubes (CNTs)
through electrostatic doping. We find that quasiparticle (QP) band gaps and
exciton binding energies can be reduced dramatically by hundreds of meVs
upon doping, and yet prominent optical absorption features shift by
relatively small amounts. Furthermore, we show that doping has a
unique influence on CNT exciton properties: in contrast to bulk excitons, 
bound excitons in semiconducting CNTs are not
quenched by doping and their binding energy can be tuned gradually even at
very high doping. These features arise due to the presence of acoustic
plasmons and their impact on dynamical screening.

We utilize a many-body \textit{ab initio} approach 
\cite{Hybertsen,Rohlfing,Spataru1} to calculate the electronic and optical
properties of electrostatically doped semiconducting CNTs. We focus on the
semiconducting (10,0) CNT, with diameter $D=0.78\mathrm{\;nm}$, and perform
\textit{ab initio} calculations \cite{note2} at zero doping and for a free carrier concentration $\rho =0.6%
\mathrm{\;holes/nm}$. We use the GW approach \cite{Hybertsen} to obtain
QP properties near the $\Gamma $ point and solve the
Bethe-Salpeter (BS) equation for the excitonic effects \cite{Rohlfing}.
Applying this approach to doped
CNTs necessitates careful consideration because of the presence of
acoustic plasmons, a unique feature of low-dimensionality
materials \cite{Hu,Lin}.

\begin{figure}[tbp]
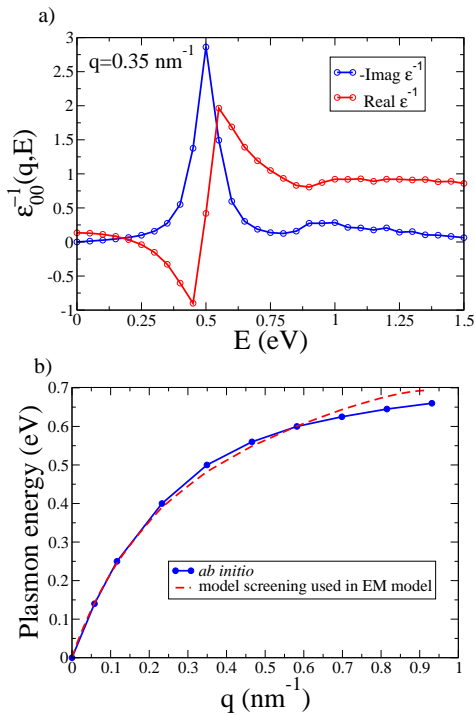

\resizebox{6.22cm}{!}{\includegraphics[clip=true]{Fig1a.eps}}
\resizebox{5.9cm}{!}{\includegraphics[clip=true]{Fig1b.eps}}
\caption{Dielectric properties of the (10,0) doped CNT. a) Real and
imaginary parts of the inverse dielectric function as a function of energy
calculated \textit{ab initio} for a fixed wave-vector $q=0.35\mathrm{\; nm}%
^{-1} $. The subscripts on $\varepsilon ^{-1}$ imply their evaluation at $\vec{G}=\vec{G\mathrm{%
^{\prime}}}=\mathrm{0}$. 
b) Plasmon dispersion relation calculated \textit{ab initio} and
within the model screening used in the effective mass model. Doping level  $\rho =0.6%
\mathrm{\;holes/nm}$.}
\label{plasmons}
\end{figure}

Indeed, in quasi-1D systems such as CNTs, electron gas and tight-binding
models predict ``acoustic'' plasmons, whose energies approach zero in the long
wavelength limit $\tilde{\omega}_{ap}(q\rightarrow 0)\propto q \sqrt{%
\rho \log (|q|D/2)}$. Our \textit{ab initio}
calculations also reveal these plasmons in doped CNTs %
\cite{truncate}: Fig. \ref{plasmons}a shows the inverse dielectric function $%
\varepsilon _{00}^{-1}(q=0.35 {\mathrm{\; nm^{-1}%
}},E)$ of the (10,0) CNT at \textit{$\rho $ =} 0.6
holes/nm. The peak in Im$%
\varepsilon ^{-1}$ signals a low-energy plasmon, which gives rise to a
transition in Re$\varepsilon ^{-1}$ between a very small value at zero
energy and a value close to 1 above the plasmon energy, \textit{i.e.}
a transition between metallic-like and semiconducting-like screening \cite%
{Spataru2}. These acoustic plasmons span a broad range of
energies, as seen in Fig. \ref{plasmons}b, and dynamical
screening effects due to them are very important for both
QPs and excitons and cannot be neglected or simply integrated out.

To include these dynamical screening effects during the $GW$ calculations we
modify methods previously applied in the context of doped bulk
semiconductors \cite{Oschlies}. In the undoped case, optical plasmons in
CNTs have energies above $\sim5$ eV \cite{Kramberger}, and their contributions 
is taken into account through the Generalized
Plasmon Pole (GPP) approximation \cite{Hybertsen}. In the doped case, GPP
alone does not describe satisfactorily dynamical effects from acoustic
plasmons and the screening is evaluated instead from $\varepsilon
^{-1}=\varepsilon _{int}^{-1}+\delta \varepsilon ^{-1}$, where $\varepsilon
_{int}$ is the dielectric function of the intrinsic semiconductor for which
we make use of the GPP approximation and $\delta \varepsilon ^{-1}$ is
obtained within the Random Phase Approximation.

Dynamical effects are included in the BS calculations by an \textit{effective%
} dielectric screening $\tilde{\varepsilon}$ that depends self-consistently
on the binding energy $E_{B}$ of the exciton \cite{Strinati}. Taking
advantage of the fact that for the excitons considered here most of the
corresponding electron-hole transitions have energies close to the onset of
the electron-hole continuum, and neglecting finite lifetime effects, one can
write to first order in dynamical effects \cite{note1}: 
\begin{equation}
\tilde{\varepsilon}^{-1}(q;E_{B})\approx \varepsilon ^{-1}(q,0)-\frac{2E_{B}}
{\pi }\int_{0}^{\infty }d\omega \frac{Im\mathrm{\;}\varepsilon
^{-1}(q,\omega )}{\omega (\omega +E_{B})}.  
\label{eps_eff}
\end{equation}
When low-energy plasmons are absent, the commonly used static approximation
is obtained by retaining the first term on the \textit{r.h.s.} of Eq. 
\eqref{eps_eff}. The second term captures dynamical effects due to
acoustic plasmons (we neglect dynamical effects due to optical plasmons
during the BS calculations in both undoped and doped cases). Smaller, higher
order corrections in dynamical effects (not shown in Eq. $\left( \ref%
{eps_eff}\right) $) due to acoustic plasmons are included as well in our
calculations.

\begin{figure}[tbp]
\resizebox{7.8cm}{!}{\includegraphics[clip=true]{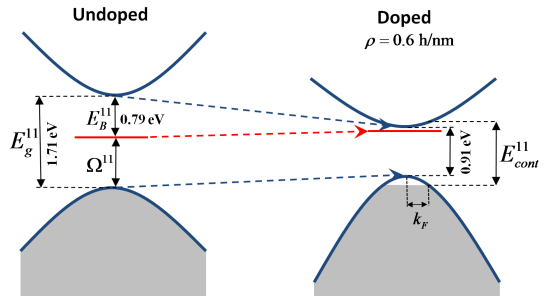}}
\caption{Illustration of bandgap and exciton renormalization in the (10,0)
CNT. The blue lines indicate the $E_{11}$ valence and conduction bands, and
the bound exciton is indicated by a red line. Occupied electronic states are
indicated by the shaded areas, and the highest occupied energy levels have
been aligned between the doped and undoped cases. The different band
curvatures represent the change in effective mass upon doping.}
\label{sketch}
\end{figure}

Fig. \ref{sketch} shows a sketch of the calculated QP bands and
exciton level associated with the lowest prominent optical transition 
($E_{11}$), before and after doping, where the energy
scale has been preserved between the two cases. The quantities of interest
are the bandgap $E_{g}^{11}$ and the exciton binding energy $E_{B}^{11}$. It
is clear from this figure that the bandgaps and exciton binding
energies are significantly reduced by doping. In fact, at this doping ($\rho =0.6%
\mathrm{\;holes/nm}$), $E_{g}^{11}$ is reduced by 800 meV, while $E_{B}^{11}$ is reduced
by 590 meV. Both of these changes are large by any measure; in particular
we estimate that band gap renormalization (BGR) is about an order of magnitude larger than in typical
bulk semiconductors at the same doping \cite{Abram}. The QP bands and exciton level associated with the second lowest prominent optical transition 
($E_{22}$, not shown in Fig. \ref{sketch}) also suffer from a significant bandgap
reduction of 130 meV \cite{note3}, and a decrease of the exciton binding energy by 240
meV. Moreover, doping leads to an important 40\% reduction in the effective
mass at the valence band maximum for $E_{11}$.

The excitonic effects are presented in more detail in Fig. \ref%
{absorb}, where the optical absorption spectra for light polarized along the
tube axis is calculated in two ways. The blue lines are obtained neglecting
the electron-hole interaction, with the corresponding onset energies
indicating the electron-hole continua. The red lines show the correct
absorption spectra obtained with the electron-hole interaction included. As
each bound exciton requires a separate self-consistent calculation in the
doped case, we have focused on the lowest energy bright exciton associated
with each of the $E_{11}$ and $E_{22}$ continua.

\begin{figure}[tbp]
\resizebox{9.0cm}{!}{\includegraphics{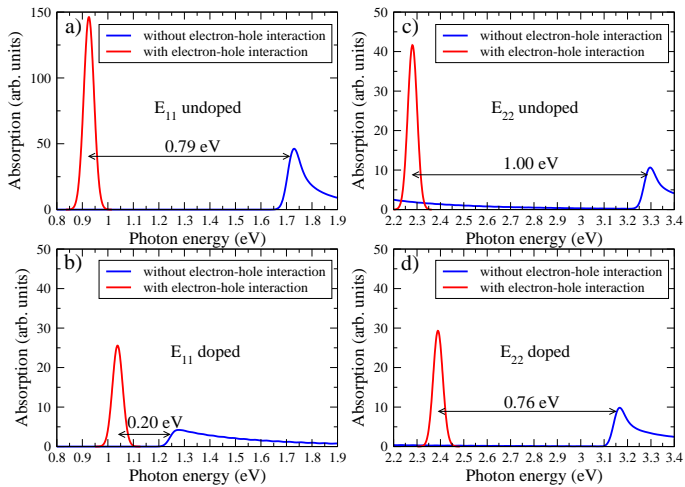}}
\caption{Impact of doping on the electronic and optical properties of the
(10,0) CNT. a,b) Optical absorption spectra near the $E_{11}$ transition
with (red) and without (blue) the electron-hole interaction. a) Before
doping. b) After doping with 0.6 holes/nm. c,d) Optical absorption spectra
near the $E_{22}$ transition c) Before doping. d) After doping with 0.6
holes/nm. Arrows indicate exciton binding energies. Optical spectra are
broadened by a 20 meV Lorentzian.}
\label{absorb}
\end{figure}

In the undoped case, the $E_{11}$ and $E_{22}$ excitons show very large
binding energies: $E_{B}^{11} (\rho =0)=0.79\mathrm{\; eV}$ and $E_{B}^{22}
(\rho =0)=1.00\mathrm{\; eV}$. As discussed above, upon
doping, a dramatic change in excitonic properties occurs. As seen in Fig. %
\ref{absorb}b, $E_B^{11}$ suffers a decrease of $%
\sim 0.6$ eV to $E_{B}^{11} (\rho =0.6\mathrm{\; holes/nm})=0.20\mathrm{\; eV%
}$, while the corresponding $E_{B}^{22} $ renormalization is 0.24 eV (see Fig. %
\ref{absorb}d). While
both the $E_{11}$ and $E_{22}$ excitons are affected by the change in
dielectric screening, the $E_{11}$ exciton renormalizes more because it is
affected by the bleaching of transitions. We also note from Fig. \ref{absorb}%
b a six-fold reduction in the oscillator strength of the $E_{11}$ exciton,
in good accord with recent photo-luminescence measurements \cite%
{Steiner} which assigned a factor of five in the drop of the $E_{11}$
exciton radiative decay rate of a 1.4 nm diameter CNT [for an estimated
maximum doping $\rho _{\max }^{\exp } \approx 0.16\mathrm{\;
holes(electrons)/nm}$].

Our \textit{ab initio} results suggest large changes in bandgap and exciton
properties upon doping \cite{note4}. This could lead, for example, to engineering of CNT
optoelectronic devices by electrostatic control; but taking advantage of
these new features requires robust control of doping-induced properties.
Because our many-body \textit{ab initio} calculations are extremely
demanding, exploring the tailoring over a broad range of doping is not
possible. Therefore, we developed a compact model for excitons in doped CNTs
based on an effective mass (EM) approach \cite{Sham}.

Our EM model seeks to describe the binding energy and envelope function of
excitons in CNTs taking advantage of their large spatial extent along the
tube axis relative to inter-atomic distances. The effective interaction
between the electron and hole composing an exciton depends on
$\tilde{\varepsilon}$ via [see
Eq. \eqref{eps_eff}] the dielectric function $\varepsilon =1-vP$, where 
\textit{P} is the irreducible polarizability. Our EM
approximation assumes that $P(z,z^{\prime},\rho ,\rho
^{\prime},\varphi ,\varphi ^{\prime})$ is localized on a cylindrical tubule
with radius \textit{R}, and that local field effects along the tube axis can
be neglected. Noting that only the \textit{L=}0 component of the angular
momentum \textit{L} is relevant for screening effects in excitons composed
of electron-hole transitions between bands of same \textit{L}, one replaces $%
P(z,z^{\prime},\rho ,\rho ^{\prime},L=0)$ with $\hat{P}(z-z^{\prime},R,L=0)$%
. Relevant dynamic effects are included via $\hat{P}(\omega) \approx \hat{P}^{free}(\omega) +\hat{P}^{int}(\omega=0) $, where $\hat{P}^{free} $  describes
intraband transitions within parabolic band approximation \cite{Hu}, and
$\hat{P}^{int} $ represents the interband transitions for the intrinsic semiconductor.
(We keep transitions from the top of valence band to
conduction states as we found that they make very small contributions.) 
$\hat{P}^{int}$ is extracted from \textit{ab initio }%
calculations for the intrinsic polarizability ($P^{int} $) by imposing that $\hat{P}%
^{int} $ and $P^{int} $ yield the same average along the radial direction over one unit
cell. In Fourier space this reads: $\hat{P}_{\mathrm{L}=\mathrm{0}}^{\mathrm{%
int}} \mathrm{(q)}=\frac{\mathrm{A}_{\mathrm{uc}} }{\pi D} P_{\vec{G}=\vec{G\mathrm{%
^{\prime}}}=\mathrm{0}}^{\mathrm{int}} \mathrm{(q)}$, where $A_{uc}$ is the
cross-sectional area of the unit cell considered in the \textit{ab initio}
case. The ability of our model -free of any adjustable parameters- in
describing dynamical screening effects from acoustic plasmons is
demonstrated in Fig. \ref{plasmons}b.

\begin{figure}[tbp]
\resizebox{9.0cm}{!}{\includegraphics{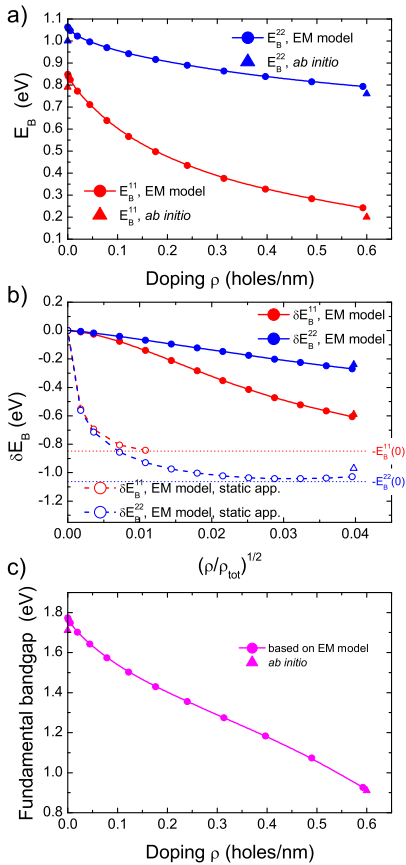}}
\caption{Exciton binding energies and fundamental bandgap as a function of
doping in the (10,0) CNT. a) Binding energies for the $E_{11}$ and $E_{22}$
lowest bright excitons calculated from the EM model and compared to 
\textit{ab initio} results. b) Change in exciton binding energies, 
$\delta E_B^S \equiv E_B^S (\rho )-E_B^S (0)$, as a function of
$\sqrt{\rho / \rho_{tot}}$ 
where $\rho _{tot}$ is the total electron 
charge density, within the EM
model with and without dynamical effects. The dotted lines refer to the EM model. The triangles represent {\textit ab initio} results: within the static approximation, we find no bound exciton associated with $E_{11}$ and $E_B^{22}\approx 30$ meV at $\rho=0.6$ holes/nm. c) Dependence of the fundamental
bandgap on doping, based on the EM model and compared to \textit{ab
initio} results.}
\label{EM}
\end{figure}

Fig. \ref{EM}a shows the EM results for $E_{B}^{11}$ and $E_{B}^{22}$
extended to carrier densities as small as $\rho \approx 1$ hole/800 nm (in
terms of number of holes per atom, this corresponds to $%
10^{17} -10^{18} \mathrm{\; holes/cm}^{3} $ in typical bulk semiconductors, 
\textit{i.e.}, approximately the degenerate limit). Comparison with \textit{%
ab initio }results illustrates the high accuracy of our EM model. Fig. \ref%
{EM}b shows that with good approximation, $\delta E_{B}^{22} \equiv
E_{B}^{22} (\rho )-E_{B}^{22} (0)\propto \sqrt{\rho } $, which we found to be 
a signature of acoustic plasmons. A similar trend is found for $\delta E_{B}^{11} $,
where deviations from the $\sqrt{\rho}$ behavior are more pronounced due to
bleaching of transitions. More importantly, the mild dependence on doping
should be contrasted to that in higher dimensional semiconductors where, at
equivalent doping levels, excitons are either quenched \cite{Schweizer} or
the dependence on doping is exponential \cite{Gubarev}, giving poor control
over optical properties.

The large change in exciton binding energy combined with the mild dependence
on doping implies that excitonic properties in CNTs can be efficiently
controlled through doping. The origin of this feature lies in the presence of acoustic
plasmons. To emphasize this point, Fig. \ref{EM}b also shows EM results for
exciton binding energies within the static approximation, \textit{%
i.e.} without proper inclusion of dynamical effects due to acoustic
plasmons. In this case, binding energies drop exponentially with doping,
much like is observed in two-dimensional materials \cite{Gubarev}; moreover,
the $E_{11}$ exciton gets quenched beyond $\rho \approx 1$hole/15 nm. The
importance of acoustic plasmons is readily seen within our model and from
Eq. \eqref{eps_eff}, which can be shown to yield: $\tilde{\mathrm{%
\varepsilon }}^{-1} \mathrm{(}q\mathrm{;E}_{B} \mathrm{)}\approx \mathrm{%
\varepsilon }^{\mathrm{-1}} \mathrm{(}q\mathrm{,0)}+\frac{E_{B} }{\tilde{%
\omega }_{ap} (q)+E_{B} } \left[\mathrm{\varepsilon }_{\mathrm{int}}^{%
\mathrm{-1}} \mathrm{(}q\mathrm{,0)-\varepsilon }^{\mathrm{-1}} \mathrm{(}q%
\mathrm{,0)}\right]$. For long wavelengths where $\tilde{\omega }_{ap}
(q)<<E_{B} $, one has $\tilde{\varepsilon }(q;E_{B} )\approx \varepsilon
_{int} (q,0)$, as opposed to the static approximation result $\tilde{%
\varepsilon }(q;E_{B} )\approx \varepsilon (q,0)$. The difference between
these two values can be orders of magnitude depending on \textit{q}.

We obtain the QP fundamental bandgap versus doping using our
exciton EM model and our many-body \textit{ab initio} results for optical
properties. Indeed, we can write $E_{g}^{11} =E_{B}^{11} +\Omega ^{11}
-\Delta E_{F}^{11} $ where $\Omega ^{11} $ is the exciton excitation energy
and $\Delta E_{F}^{11} \equiv E_{cont}^{11} -E_{g}^{11} $ (see Fig. \ref{sketch}). 
We calculate the doping dependence of $\Delta E_{F}^{11} $ from $\Delta
E_{F}^{11} \cong k_{F}^{2} /2\mu ^{*} $, with the reduced exciton mass $\mu
^{*} $ obtained at various doping levels by interpolating the values
from many-body \textit{ab initio} calculations at $\rho =0$ and $%
\rho =0.6\mathrm{\; holes/nm}$. Furthermore, we find from our \textit{ab
initio} calculations at $\rho =0.6\mathrm{\; holes/nm}$ that $\Omega ^{11} $
only increases by $\sim 0.1$ eV upon doping (similarly for $\Omega ^{22} $), a result of cancelation between 
self-energy corrections and excitonic effects \cite%
{Burstein}. The smallness of these shifts is in excellent agreement with
measurements of the $E_{33} $ absorption peak of a 1.4 nm
diameter CNT \cite{Steiner}, showing a red-shift of 20 meV at $\rho _{\max
}^{\exp }$. Therefore, with little expected error, we assume a linear
dependence of $\Omega ^{11} $ on doping, and plot in Fig. \ref{EM}c the
fundamental bandgap versus doping; the trend indicates that BGR 
can also be tuned gradually over a broad energy range.

The giant BGR discussed here is in 
agreement with recent experimental results \cite{Lee} that combined
photocurrent spectroscopy with transport measurements:
for a 1.5 nm diameter CNT with QP band gap of 0.91 eV at zero doping, a BGR
of 0.54 eV was deduced at a doping density of 0.7 electrons/nm.

In conclusion, we have shown that doping has a profound and unique impact on
the electronic and optical properties of semiconducting CNTs, and that
dynamical effects from acoustic plasmons are essential to capture these
effects. Our study indicates that bandgaps and exciton binding energies in
CNTs can be tuned significantly and gradually by electrostatic doping,
establishing a new framework for the understanding and design of
CNT-based devices. We expect that
similar control will be possible in a broad range of nanomaterials.

Sandia is a multiprogram laboratory operated by Sandia Corporation, a
Lockheed Martin Company, for the United States Department of Energy's
National Nuclear Security Administration under Contract DE-AC04-94AL85000.
Work supported by the Lockheed Martin Shared Vision program.

\noindent * Corresponding author. E-mail: cdspata@sandia.gov

\clearpage


\begin{references}
\bibitem{note1}
We stress that $\tilde{\varepsilon }^{-1} (q;E_B )\ne \varepsilon ^{-1} (q,\omega =E_B )$, as an electron excited in a conduction band probes the density response of the system with a hole in the valence band \cite{Strinati}.
\bibitem{note3}
Within GPP alone during GW, $E_g^{11}$ and $E_g^{22}$ are reduced upon doping by $890$ meV and $800$ meV respectively.
\bibitem{note4}
Upon doping with 0.6 holes/nm, the $E_{11}$ exciton delocalizes over tens of nm along the tube axis (at zero temperature, the envelope function fall-off at large electron-hole separation is algebraic in the doped case, in contrast to the exponential decay in the undoped case) and acquires nodes  separated by $\sim 5$ nm.
\end{references}
\end{document}